\documentclass[a4paper,12pt]{article}
\usepackage{cite,amssymb}

\advance\textwidth 0.5in \advance\oddsidemargin -0.25in \advance\textheight 0.5in \advance\topmargin -0.5in

\def\bea{\begin{eqnarray}}
\def\eea{\end{eqnarray}}

   \def\({\left(} \def\){\right)}

\def\beq{\begin{equation}} \def\eeq{\end{equation}} \def\ba{\begin{array}} \def\ea{\end{array}}

\title {\bf On a nonlinear integrable difference equation on the square}

\author{{\bf D. Levi} \\ Dipartimento di Ingegneria Elettronica, \\ Universit\`a degli Studi Roma Tre and Sezione INFN, Roma Tre, \\ Via della
Vasca Navale 84, 00146 Roma, Italy \\ {\sl E-mail: levi@roma3.infn.it} \and {\bf R.I. Yamilov} \\ Ufa Institute of Mathematics, Russian Academy
of Sciences, \\ 112 Chernyshevsky Street, Ufa 450077, Russian Federation \\ {\sl E-mail: RvlYamilov@matem.anrb.ru}}

\date{\today} 
\begin{document} \maketitle

\begin{abstract} We present a nonlinear partial difference equation defined on a square which is obtained by combining the Miura transformations between the Volter\-ra and the modified Volterra differential-difference equations. This equation is not symmetric with respect to the exchange of the two discrete variables. Its integrability is proved by constructing its Lax pair.  \end{abstract}


The uncovery of new nonlinear integrable completely discrete equations is always a very challenging problem as, by proper continuous
limits, many other results on
differential-difference and partial differential equations can be obtained. In
the case of differential  equations by now a lot is known starting from the
pioneering works by Gardner, Green, Kruskal and Miura. A summary of these
results is already of public domain and presented for example in the
Encyclopedia of Mathematical Physics \cite{emp} or in the Encyclopedia of Nonlinear Science \cite{ens}. Among those results let
us mention the classification scheme of nonlinear integrable partial
differential equations introduced by Shabat using the formal symmetry
approach, see \cite{msy} for a review. The classification of
differential-difference equations has also been carried out using the formal
symmetry approach by Yamilov \cite{y83} and it is a well defined procedure which can be easily computerized for many
families of equations \cite{ly,y06}. 

In the completely discrete case the situation
is different. Many researchers have tried to carry out the approach of formal
symmetries introduced by Shabat, without any success up to now.  One of the first
exhaustive results in this context, based on completely different ideas,
is given by the Adler-Bobenko-Suris (ABS)
classification
of $\mathbb{Z}^2$-lattice equations defined on the square lattice
\cite{abs}. By now many results are known
on the ABS equations, see for instance \cite{ra,gr,lp,lps}. However the
analysis of the transformation properties of these lattice equations cannot be considered yet complete
and new results which help the understanding of the interrelations between
them and some differential-difference equations can still be found \cite{lpsy}.

A two-dimensional partial difference equation is a functional relation among the values of a function
$u: \mathbb{Z} \times \mathbb Z \rightarrow \mathbb C$ at different points of
the lattice of indices $i,j$. It involves the independent
variables $i,j$ and the lattice parameters $\alpha, \beta \in \mathbb C$:
$$
\mathcal E (i,j, u_{i,j}, u_{i+1,j},u_{i,j+1},...; \alpha, \beta)=0.
$$

The so-called ABS list of  integrable lattice equations is given by those affine linear 
(i.e. polynomial of degree one in each argument) partial difference
equations of the form
\beq\label{jj}
\mathcal E (i,j, u_{i,j}, u_{i+1,j}, u_{i,j+1}, u_{i+1,j+1}; \alpha, \beta)=0, 
\eeq
whose integrability is based on the {\it consistency around a cube} (or 3D-consistency) \cite{abs}. 

\begin{figure}[htbp]
\begin{center}
\setlength{\unitlength}{0.08em}
\begin{picture}(200,140)(-50,-20)
  \put( 0,  0){\line(1,0){100}}
  \put( 0,100){\line(1,0){100}}
  \put(  0, 0){\line(0,1){100}}
  \put(100, 0){\line(0,1){100}}
  \put(-32,-13){$u_{i,j}$}
    \put(-19,47){$\beta$}
     \put(47,-15){$\alpha$}
  \put(103,-13){$u_{i+1,j}$}
  \put(103,110){$u_{i+1,j+1}$}
  \put(-32,110){$u_{i,j+1}$}
\end{picture}
\caption{A square lattice}
\end{center}
\end{figure}
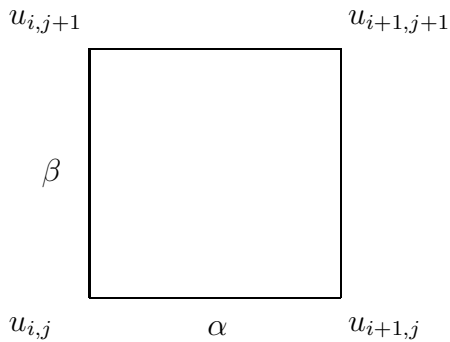

\begin{figure}[htbp]
\begin{center}
\setlength{\unitlength}{0.08em}
\begin{picture}(200,170)(-50,-20)
  \put(100,  0){\circle*{6}} \put(0  ,100){\circle*{6}}
  \put( 50, 30){\circle*{6}} \put(150,130){\circle*{6}}
  \put(  0,  0){\circle{6}}  \put(100,100){\circle{6}}
  \put( 50,130){\circle{6}}  \put(150, 30){\circle{6}}
  \put( 0,  0){\line(1,0){100}}
  \put( 0,100){\line(1,0){100}}
  \put(50,130){\line(1,0){100}}
  \multiput(50,30)(20,0){5}{\line(1,0){15}}
  \put(  0, 0){\line(0,1){100}}
  \put(100, 0){\line(0,1){100}}
  \put(150,30){\line(0,1){100}}
  \multiput(50,30)(0,20){5}{\line(0,1){15}}
  \put(  0,100){\line(5,3){50}}
  \put(100,100){\line(5,3){50}}
  \put(100,  0){\line(5,3){50}}
  \multiput(50,30)(-16.67,-10){3}{\line(-5,-3){12}}
     \put(-10,-13){$u_{i,j,k}$}
     \put(90,-13){$u_{i+1,j,k}$}
     \put(50,17){$u_{i,j,k+1}$}
     \put(-13,110){$u_{i,j+1,k}$}
     \put(160,25){$u_{i+1,j,k+1}$}
     \put(45,140){$u_{i,j+1,k+1}$}
     \put(109,95){$u_{i+1,j+1,k}$}
     \put(157,135){$u_{i+1,j+1,k+1}$}
     \put(40,-13){$\alpha$}
     \put(-16,50){$\beta$}
     \put(20,25){$\gamma$}
\end{picture}
\caption{Three-dimensional consistency}\label{fig.cube}
\end{center}
\end{figure}
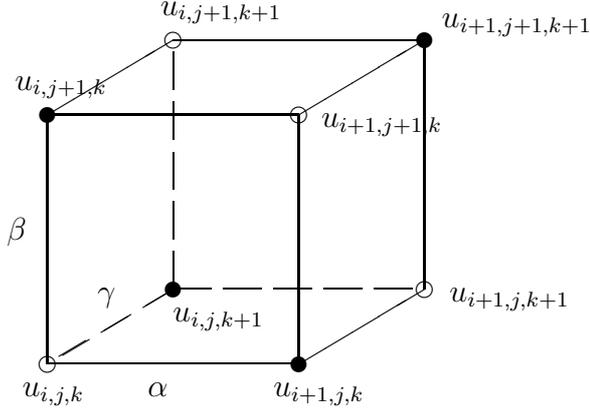

The main idea of the consistency method is the
following.
One starts from a square lattice, defines the variables on the
vertices $u_{i,j}, u_{i+1,j}, u_{i,j+1}, u_{i+1,j+1}$ (see Figure 1)
and considers the multilinear equation relating these variables, namely eq. (\ref{jj}). 
By solving it for $u_{i+1,j+1}$ one obtains a rational expression and the same holds for any field variable. One then
adjoins a third direction and imagines the map giving $u_{i+1,j+1,k+1}$ as being the composition of maps on the various planes (see Figure 2). There
exist three different ways to obtain $u_{i+1,j+1,k+1}$ and the consistency constraint is that they all lead to the same result. This
gives strict conditions on the nonlinear equation, but they are not sufficient
to determine it completely. 
Two further constraints have been
introduced by Adler, Bobenko and Suris. They are:

\begin{itemize}

\item {\it $D_4$-symmetry}. $\mathcal E$ is invariant under the group of the square symmetries:
\[
\mathcal E (u_{i,j}, u_{i+1,j}, u_{i,j+1}, u_{i+1,j+1}; \alpha, \beta) =
 \pm \mathcal E (u_{i,j}, u_{i,j+1}, u_{i+1,j}, u_{i+1,j+1}; \beta, \alpha)  =
\]\[
  \pm \mathcal E (u_{i+1,j}, u_{i,j}, u_{i+1,j+1}, u_{i,j+1}; \alpha, \beta).
\]
 
\item {\it Tetrahedron property}. The function $u_{i+1,j+1,k+1}$ is independent of $u_{i,j,k}$.

\end{itemize}

The following transformations, which do not violate the two constraints listed above, are assumed to identify equivalence classes:

\begin{itemize}
\item Action on all field variables by one and the same (independent of lattice parameter) M\"obius transformation.
\item Simultaneous point change of all parameters.
\end{itemize}

Under the above constraints Adler, Bobenko and Suris obtained a complete classification of $\mathbb{Z}^2$-lattice systems, whose
integrability is ensured as the {\it consistency around a cube} also furnishes their Lax pairs \cite{abs,bob,ni}.   

As it is known \cite{lss}, the modified Volterra equation \beq\label{b1} u_{i,t} = (u_i^2 -1) (u_{i+1} - u_{i-1}) \eeq is transformed into the Volterra
equation $v_{i,t} = v_i (v_{i+1} - v_{i-1})$ by two discrete Miura transformations: \beq\label{b2} v_i^\pm = (u_{i+1} \pm 1) (u_i \mp 1) . \eeq
For any solution $u_i$ of eq. (\ref{b1}), one obtains by the transformations (\ref{b2}) two solutions $v_i^+,v_i^-$ of the Volterra equation. From a solution of the Volterra equation $v_i$ we obtain two solutions of the modified Volterra equation $u_{i,0}$ and $u_{i,1}$. The
composition of the  Miura transformations (\ref{b2}) 
\beq\label{b3} 
v_i = (u_{i+1,0} +1) (u_{i,0} -1) = (u_{i+1,1} -1) (u_{i,1} +1) 
\eeq 
provides a B\"acklund transformation for eq. (\ref{b1}). Eq. (\ref{b3}) provides a way  to construct from a solution $u_{i,0}$ of eq. (\ref{b1}) a new solution
$u_{i,1}$. Iterating  eq. (\ref{b3}),  one can construct infinitely many solutions:
\[
\dots \leftarrow  u_{i,-2} \leftarrow  u_{i,-1} \leftarrow  u_{i,0} \rightarrow u_{i,1} \rightarrow u_{i,2} \rightarrow \dots 
\]
Rewriting eq. (\ref{b3}) as a chain of equations relating the solutions $u_{i,j}$,
we obtain the following completely discrete equation on the square:
\beq\label{b4} 
(u_{i+1,j} +1) (u_{i,j} -1) = (u_{i+1,j+1} -1) (u_{i,j+1} +1) . 
\eeq

This equation does not belong to the ABS classification, as it is not invariant under the exchange of $i$ and $j$. However  eq. (\ref{b4}) is invariant under a rotational symmetry of $\pi$.  By a straightforward calculation, using a symbolic computation program like Maple, one can easily show its 3D-inconsistency. 
Recently Adler, Bobenko and Suris \cite{abs1} extended the previous definition to  systems of equations 3D-consistent on a cube to the case when the two equations of the Lax pair are different. Then eq. (\ref{b4}) can be embedded into such a 3D-consistent system \cite{tx}. Moreover eq. (\ref{b4}) can be easily transformed in the discrete version of the Volterra--Kac--van Moerbeke equation \cite{NC}.

The construction of the Lax pair can be done in a way that is parallel to the derivation of the nonlinear difference equation done above. Let us consider the spectral problem for the modified Volterra equation (\ref{b1}) 
\bea \label{l1}
L_i = 
\left( \begin{array}{cc} 
-\lambda^{-1} &  u_i \\
-u_i & \lambda 
\end{array} \right) ,
\eea
found in \cite{al}, and the standard scalar spectral problem of the Volterra equation, written in matrix form,
\bea \label{l2}
M_i = 
\left( \begin{array}{cc} 
\lambda -\lambda^{-1} &  -v_i \\
1 & 0 
\end{array} \right).
\eea
The existence of the two Miura transformations (\ref{b2}) between the two equations imply the existence of two nonsingular Darboux matrices $E^{(+)}_{i}, E^{(-)}_{i}$ between the spectral problems:
\bea \label{l3}
E^{(+)}_{i} = 
\left( \begin{array}{cc} 
1 & \lambda v_i(u_{i,0} + 1) \\
\lambda & - v_i(1 + u_{i,0})
\end{array} \right), \qquad 
E^{(-)}_{i} = 
\left( \begin{array}{cc} 
-1 & \lambda v_i(u_{i,1} -1) \\
\lambda & - v_i(1 -u_{i,1})
\end{array} \right) .
\eea
The matrix $E^{(+)}_{i}$ will provide a solution $u_{i,0}$ of the modified Volterra equation, while the matrix $E^{(-)}_{i}$ will provide a different solution, $u_{i,1}$. So, the two solutions $u_{i,0}$ and $u_{i,1}$ are given by the two Lax equations
\bea \label{l4}
E^{(+)}_{i+1} M_i = L_{i,0} E^{(+)}_{i} , \qquad E^{(-)}_{i+1} M_i = L_{i,1} E^{(-)}_{i} ,
\eea
where
\bea \label{ll}
L_{i,j} = 
\left( \begin{array}{cc} 
-\lambda^{-1} &  u_{i,j} \\
-u_{i,j} & \lambda 
\end{array} \right) .
\eea

The equation (\ref{b3}), relating the two solutions  $u_{i,0}$ and $u_{i,1}$, is obtained by eliminating from eqs. (\ref{l4}) the matrix $M_i$ and the dependence of $v_i$. So its Lax equation is given by 
\bea \label{l5}
N_{i+1,0}  L_{i,0} = L_{i,1} N_{i,0} ,
\eea
where $N_{i,0} =  E^{(-)}_{i} ( E^{(+)}_{i})^{-1}$. Taking into account the definition (\ref{l3}), formulae (\ref{b3}) for $v_i$, the discrete equation (\ref{b4}), and introducing as before the chain of equations for any $j$, we get that the Lax equation associated to eq. (\ref{b4}) is given by 
\[ 
N_{i+1,j}  L_{i,j} = L_{i,j+1} N_{i,j},
\]
with $L_{i,j}$ given by eq. (\ref{ll}) and 
\[
N_{i,j} =  
\left( \begin{array}{cc} 
\lambda w_{i,j} - \lambda^{-1} & -(w_{i,j} + 1) \\
w_{i,j} + 1 & \lambda - \lambda^{-1} w_{i,j} 
\end{array} \right) , \qquad 
w_{i,j} = \frac{u_{i,j} +1}{u_{i,j+1} -1} .
\]

This is not the only case when we can encounter 3D--inconsistent integrable equations. For example,  the modified--modified Volterra equation will provide in the same way a discrete equation on the square
\bea \label{a16}
 (1 + u_{i,j} u_{i+1,j})(\mu u_{i+1,j+1} + \mu^{-1} u_{i,j+1})=(1 + u_{i,j+1} u_{i+1,j+1})(\mu u_{i,j} + \mu^{-1} u_{i+1,j}), 
\eea
where $\mu$ is an arbitrary non--zero constant.
This equation has the same symmetry properties as eq. (\ref{b4}) and is also 3D--inconsistent when $\mu^4 \ne 1$. For $\mu^4=1$ eq. (\ref{a16}) is 3D--consistent, but in this case the equation is degenerate and can be written as $(T_j \pm 1 ) \frac{\mu u_{i,j}+ \mu^{-1} u_{i+1,j}}{1 + u_{i,j} u_{i+1,j}} =0$, where $T_j$ is the shift operator for the  $j$ index. Also eq. (\ref{a16}) can be embedded into a  system 3D-consistent on a cube \cite{tx}.

\paragraph{Acknowledgments.} R.I.Y. has been partially supported by the Russian Foundation for Basic Research (Grant numbers 07-01-00081-a and 08-01-00440-a). D.L. has been partially supported by PRIN Project {\it Metodi matematici nella teoria delle onde nonlineari ed applicazioni -- 2006} of the Italian Ministry of Education and Scientific Research. R.I.Y. and D.L. thank the Isaac Newton Institute for Mathematical Sciences for their hospitality during the {\it Discrete Integrable Systems} program and thank A. Tongas and P. Xenitidis for useful discussions.

 \end{document}